# Techniques for Fast Transient Fault Grading Based on Autonomous Emulation

Celia López-Ongil, Mario García-Valderas, Marta Portela-García, Luis Entrena-Arrontes
Microelectronics Group. Electronic Technology Department
University Carlos III of Madrid. Spain
celia, mgvalder, mportela, entrena@ing.uc3m.es

**Abstract**
*Very deep submicron and nanometer technologies have increased notably integrated circuit (IC) sensitiveness to radiation. Soft errors are currently appearing into ICs working at earth surface. Hardened circuits are currently required in many applications where Fault Tolerance (FT) was not a requirement in the very near past. The use of platform FPGAs for the emulation of single-event upset effects (SEU) is gaining attention in order to speed up the FT evaluation. In this work, a new emulation system for FT evaluation with respect to SEU effects is proposed, providing shorter evaluation times by performing all the evaluation process in the FPGA and avoiding emulator-host communication bottlenecks.*

## I. INTRODUCTION

With the advent of very deep submicron technologies fault tolerance (FT) is becoming a concern for an increasing number of applications, since they can be affected by radiation even in the earth surface. During the hardening process of a circuit FT evaluation is a key factor. The injection of real faults artificially generated in a circuit prototype [1] is a hardware approach for this task. It produces very realistic results, but requires expensive equipment. Also, identification of weak areas is difficult in the circuit prototype, due to the limited observability at the chip pins. So, re-design cost increases notably.

Software approach for FT evaluation is modelling the behaviour of transient faults. Fault model is injected in a circuit description and validation of the faulty circuit response, by means of simulation, will indicate the robustness of the design. In fault simulation re-design for hardening could be easily done; early location of weak areas will imply a significant reduction in re-design cost and time. However the flexibility achieved in the measuring of the FT implies a great cost in terms of CPU time and resources.

Simulation of large designs is currently being substituted by hardware emulation in platform FPGAs that provides a considerable saving of computational resources. Emulation is controlled by a host computer that interacts with the FPGA hardware. This way, software flexibility is maintained while hardware speed is profited. The application of emulation to FT evaluation has been proved to be a very cost effective solution. It is important to note that in fault emulation only functional response is being tested, not physical implementation response.

In this paper, we proposed a new approach for performing FT evaluation by means of autonomous emulation of the circuit in platform FPGAs. This solution makes profit of the hardware resources, executing most of the task in the FPGA instead of in the host, in order to minimise the bottleneck times in the communication between software and hardware. Three different techniques for autonomous emulation are presented and compared, stating that best technique depends on the characteristics of the circuit. This work was partially supported by research project nb. 07T/0052/2003 2.

## II. AUTONOMOUS EMULATION SYSTEM

Hardware emulation in FT evaluation consists in testing the response of a faulty circuit, which has been modified in order to behave according to a fault model. Commonly, *bit-flip* is the fault model adopted for SEU effects. Therefore, only memory elements are affected.

The fastest existing emulation solutions for FT evaluation are based on the modification of the circuit under evaluation adding extra hardware to allow fault injection from a host computer [2]. The main drawback of this approach is the host-emulator communication bottleneck, because it is required for multiple tasks, like stimuli application, fault injection and output values check.

The proposed autonomous emulation system solves this problem by performing all the required tasks in the FPGA. The system includes a controller to manage all the emulation tasks and a RAM to store inputs, outputs and fault classification. The evaluation process can be performed in a single hardware run, reducing emulator-host communication to the beginning and to the end of the whole process.

Three fault injection techniques have been implemented within the proposed emulation system. The *mask-scan* technique departs from [2], adding the required components for the system to become autonomous. The *state-scan* technique and the *time-multiplexed* technique have been originally developed in this work.

In the *mask-scan* technique, an additional flip-flop is added to every circuit flip-flop to be used as a fault mask. Mask flip-flops indicate where a fault is to be injected.

In the *state-scan* technique, circuit flip-flops form a scan chain that allows the insertion of the whole faulty circuit state. These states are stored in RAM and the emulation controller can retrieve and insert them in the circuit.

Finally, the *time-multiplexed* technique replaces every circuit flip-flop by the structure shown in Figure 1. In this structure, two flip-flops are used to run the faulty circuit and the golden run alternatively, thus performing time multiplexing. A third flip-flop is used as a fault mask, and a state flip-flop is also used to avoid restarting the emulation from the beginning every time. This technique is quite faster than the previous ones, because it allows detecting fault effects disappearing without executing the whole testbench.

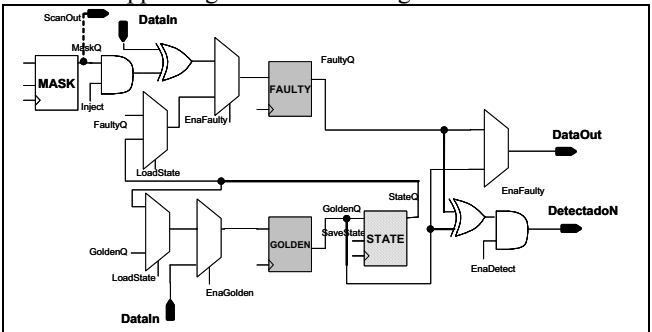

**Figure 1. Instrument for the time multiplexed technique**

## III. EXPERIMENTAL RESULTS AND CONCLUSION

The autonomous emulation system has been implemented in a Celoxica RC1000 board, with a Xilinx Virtex-2000E and 8Mb of onboard RAM.

The b14 circuit from the ITC'99 benchmark suite [3], the Viper processor, has been used. The circuit has 32 inputs, 54 outputs and 215 flip-flops. The experiments have been carried out with a set of 160 stimulus vectors and the complete set of single faults. Synthesis results for the original circuit, the modified versions and the complete emulation circuit are shown in table 1(Leonardo Spectrum 2003).

|  | Board /FPGA RAM | Modified circuit (overhead) | | Emulator System (overhead) | |
| --- | --- | --- | --- | --- | --- |
|  |  | LUTs | FFs | LUTs | FFs |
| **b14 original** | - | 1,172 | 215 | - | - |
| **Mask Scan** | 33 / 13.4 kbits | 1,657 (41%) | 434 (102%) | 2,040 (74%) | 670 (211%) |
| **State Scan** | 7,289 / 13.4 kbits | 1,644 (40%) | 433 (101%) | 1,728 (47%) | 518 (140%) |
| **Time Multiplex.** | 67 / 5.3 kbits | 3,836 (227%) | 859 (300%) | 4,162 (255%) | 1,032 (380%) |

**Table 1. Synthesis results for the b14 circuit**

The overhead due to circuit modification is proportional to the flip-flop number. State-scan technique also uses a higher amount of memory to store the different faulty states to evaluate. Control block overhead depends on the flip-flop number, test bench cycles and circuit inputs and outputs.

The set of 34,400 single faults have been classified into a 49.2% failure, 4.4% latent and 46.4% silent faults.

Emulation times are shown in table 2. With a clock frequency of 25 MHz the average speed obtained is some orders of magnitude better than fault simulation (1300 us/fault) and emulation in [2](100 us/fault).

| Autonomous System | Emulation time (ms) | Average speed (us/fault) |
| --- | --- | --- |
| **MaskScan** | 141.11 | 4.1 |
| **State Scan** | 386.40 | 11.2 |
| **Time Mux.** | 19.95 | 0.58 |

**Table 2. Time results for b14 circuit**

Emulation time in the *State-Scan* technique is longer due to the need of scanning-in the circuit state for each fault. This is because in this circuit (b14) there are a large number of flip-flops (215) and a short number of test bench cycles (160). This method improves when the number of cycles is higher than the flip-flop number. *Time-Multiplexed* technique is always the fastest.

Finally as a conclusion, this paper presents a new solution for improving the performance of transient fault emulation in platform FPGAs. An autonomous transient fault emulation system is proposed, which executes most of the tasks involved in the fault injection campaign in the FPGA. Experimental results have shown that the proposed system achieves execution times several orders of magnitude faster than existing solutions, with a reasonable cost in terms of extra hardware thanks to the use of platform FPGAs. A comparison has been made between three techniques that apply autonomous emulation, with respect to emulation time and area overhead.